# Smart Passive Acoustic Monitoring:
# Embedding a Classifier on AudioMoth Microcontroller

**Louis Lerbourg[1], Paul Peyret[2], Juliette Linossier[2], Marielle Malfante[1]**

[1]Univ. Grenoble Alpes, CEA, List, Grenoble, France
[2]Biophonia, France

louis.lerbourg@cea.fr, marielle.malfante@cea.fr

## Abstract

Passive Acoustic Monitoring (PAM) is an efficient and non-invasive method for surveying ecosystems at a reduced cost. Typically, autonomous recorders allow the acquisition of vast bioacoustic datasets which are then analyzed. However, power consumption and data storage are both scarce and limit the duration of acquisition campaigns. To address this issue, we propose a smart PAM system which allows the in-situ analysis of the soundscape by embedding a classifier directly onto an AudioMoth microcontroller. Specifically, we propose an optimized yet simple 1D Convolutional Neural Network (1D-CNN) to classify the raw audio. The model focuses on the specific call of Scopoli's Shearwater seabirds (endangered species) and is trained on a real-world dataset with a classification accuracy of 91% (balanced accuracy of 89%). We also propose a process to optimize the model to fit the severe resource constraints of the AudioMoth, achieving a ~10kB RAM memory footprint and 20ms inference time. Finally, we present an open-source tutorial of our model optimization and export strategy which can be used for embedding models beyond the scope of our study. Our modified version of the AudioMoth firmware adds two functions: (F1) which selectively records data when the target species has been detected and (F2) which logs the continuous classification results in real time. This work intends to facilitate the conception of intelligent sensors, enhancing the efficiency and scalability of bioacoustic monitoring campaigns.

## Introduction

Passive Acoustic Monitoring (PAM) [1], [2] uses autonomous recorders like AudioMoth [3] to collect vast quantities of acoustic data from a given environment. The collected data is later post-processed to extract relevant information on the considered environment or on specific species. Such process and tools for efficient biodiversity monitoring are crucial for conservation efforts. If the collection of raw and indiscriminate soundscape can be relevant to many studies, we here focus on the possibility to perform in-situ analysis. Typically, performing the soundscape analysis locally and in real-time can allow for more specific recording strategies, for instance by triggering a recording only if a given species is detected. It can also bypass the acquisition of the raw data by directly recording a continuous log of the soundscape therefore gaining memory storage and eventually send regular report on the studied environment or even alarms if necessary (e.g. illegal activities detection).
In the literature, AI classification on bioacoustics data mostly rely on pre-processing input into spectrograms like for BirdNET models [4], [5] and if this approach is successful it is also computationally intensive. Alternative approaches processing directly the raw waveform exist but are not widely deployed despite their interest from a computational point of view [6]. Successfully deploying neural networks on resource-constrained hardware is greatly eased using specialized frameworks that provide dedicated optimization and export tools. We mention in particular Tensorflow, PyTorch, Edge-Impulse, or more recently the framework Aidge which differs as an open-source alternative designed for verifiable and certifiable embedded systems without any opaque or proprietary modules [7], [8].
In this study we focus on a widely-used acoustic recorder: AudioMoth [3] a bare-metal microcontroller with ARM Cortex-M4 processor, microphone, FPU, 265kB FLASH, 32kB of RAM and 256kB of SRAM. Notably, the real free RAM after consideration of the user firmware is ~10kB. To the best of our knowledge, previous work embedding a classifier onto the Audiomoth are scarce: [9] embedded a classifier to detect cicadas and gunshots with a non-neuronal approach and an OpenAcousticDevice project[1] integrated Edge-Impulse neural network classifiers but lacks any reference in the literature. Recently, we can also refer to [10] with work on the MFCCs.

In this work, we develop and integrate an efficient deep learning model into the AudioMoth firmware, respecting the remaining memory budget of ~10kB. In particular, our contributions are: (1) the design of a minimal-footprint 1D-CNN classifier for Shearwater calls operating on raw audio, (2) the development into Aidge of a tiling method[2] for the optimization and export of our model, (3) the creation and demonstration of a firmware for AudioMoth with two operative modes (F1) *detection-triggered-recording* (Figure 2) and (F2) *analyse-and-record* which is also available as a video[3], (4) a quantification of the performance, latency, and energy of our smart PAM system.
This paper is associated to the demonstration of the Smart AudioMoth presented at ICASSP2026 (track demo).

## System Architecture and Results

Our model is trained on a dataset of Scopoli's Shearwater calls containing approximately 10,000 labeled events from 24kHz wav recordings sampled with 16 bits resolution. Four main classes are present: three positive classes of male, female, chicks and a negative class containing background noise. Recordings are sourced from ten different locations, using ten different recorders thereby ensuring significant environmental and acoustic variabilities. Our model architecture is composed of 12 1D-convolutional layers for feature extraction (16 filters, a dilation factor of 5 and stride of 3 every 3 convolutional layer), followed by one dense layer of size four (output after Softmax activation). The model processes audio in consecutive and standardized windows of 9000 samples (W=375ms). The model is trained using Focal Loss [11]. This first optimization process resulted in the 62kB model described in Table 1 achieving 89% balanced accuracy and 91% accuracy on the test set.

To meet the strict ~10kB RAM budget of the AudioMoth, we further optimized the model with input of 1024 samples (W=42.7ms) and an architecture of six convolutions (4 filters, stride every 2 convolutions and dilation rate of 3), resulting in the 3.3kB model of Table 1 and reaching a balanced accuracy of 79% and an accuracy of 81%. This 3.3kB model is exported to the ONNX format and converted by Aidge to a C++ source. Even with this reduced model the main challenge is the peak RAM usage during convolutions which we address by implementing a tiled convolution strategy on the convolutional layers. Typically, a demanding operation is divided into smaller and sequential blocks thereby reducing the necessary amount of RAM at any given time by: a) splitting the last layer output into multiple slices, b) attributing to each output slide its corresponding input (i.e. its receptive field), c) executing the layers sequentially on each input slice and d) concatenating the results to reconstruct the overall output of the convolutional layers. Operating on five slices allows the inference RAM peak to drop from 36.8kB to 10.2kB therefore fitting into the AudioMoth RAM budget.

| model | window (ms) | accuracy | balanced accuracy | RAM peak (kB) | RAM peak tiling (kB) |
|---|---|---|---|---|---|
| 62kB | 375 | 0.91 | 0.89 | 1184.2 | |
| 3.3kB | 42.7 | 0.81 | 0.79 | 36.8 | 10.2 |

Table 1- Models summary: 1st model reaches 91% accuracy, 2nd model fits under 10kB in RAM. Tiling reduces RAM peak allowing real-time analysis.

On board, the Direct Memory Access (DMA) controller continuously writes audio samples from the microphone into an SRAM buffer. This operates independently from the CPU which analyses the already-buffered windows. The new firmware option (F2) *analyse-and-record* runs standardization and inference on the consecutive windows. The complete execution has a duration of 36ms for a 42.7ms input window which allows for real-time analysis, in parallel of the next window acquisition. Playing on the speaker the 686 events of our test dataset showed that the system achieves a 100% accuracy on the test dataset in lab conditions. Alternatively, the function (F1) *detection-triggered-recording* runs detection cycles of 10 seconds during which the consecutive windows are standardized before inference by the model, as illustrated in Figure 1. A positive shearwater detection is triggered if the cumulative count for any single Shearwater class (male, female, or chick) surpasses a threshold of 30 positive classifications (to ensures that only persistent vocalizations trigger a detection).

A current consumption and latency analysis of the Smart-AudioMoth is made measuring voltage at terminals of a shunt 1.045Ω resistor, with results illustratedFigure 1. Notably, an analysis cycle of 42.7ms with (F2) containing pre-processing, inference and rest phases costs 8.31mJ compared to 7.03mJ on the regular AudioMoth. It corresponds to an additional 18% in energy consumption if inference is done locally.

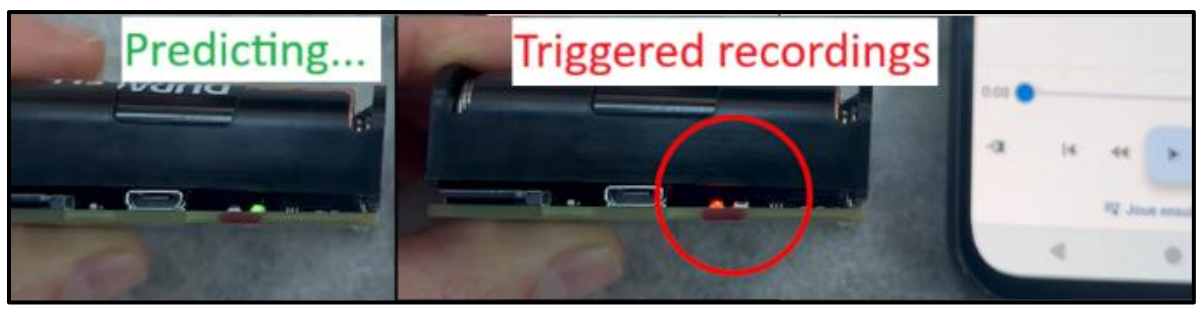

Figure 2: Demonstration of Audiomoth's triggered recording after detection of a Shearwater's call from a speaker.

## Demonstration content

The demonstration and associated video[3] illustrating this paper are show the two different functions of our system: (F1) *detection-triggered-recording* and (F2) *analyse-and-record*. The elements of analysis regarding the systems energy consumption, latency and performances are also illustrated with the experimental set-up and oscilloscope.

The function (F1) illustrated in Figure 2 shows the device correctly ignoring ambient noise and triggering a recording only a Shearwater call is played by an external source. This confirms the core functionality and the saving in both battery and storage if no species of interest is present. Visually, the green LED signals the 10-second analysis stage and, after we played a shearwater call on an audio monitor, the red LED signals that the device is now saving the recording to the SD card. The oscilloscope-based setup which lead to results presented in Figure 2allows to clearly show on computer screen the real-time continuous classification results of the soundscape (using a EFM32WG Starter Kit as a debugger), and on the oscilloscope the current consumption profile and latency of the system in sleep, analysis and recording modes.

Lastly, the Jupyter notebook tutorial[2] we made as a contribution to the Aidge platform is presented as a brief overview in the video. It highlights Aidge's C++ code generation capability and the tiling method we created.

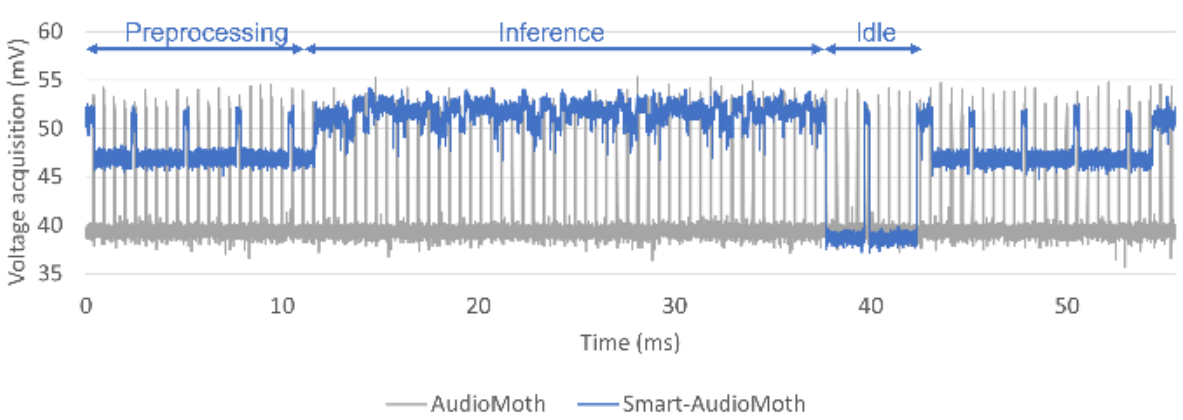

Figure 1- Voltage comparison between original AudioMoth and Smart AudioMoth running with (F2) for continuous analysis of the soundscape. Three stages are visible on the plot: preprocessing, inference and idle.

## Conclusion and Future Work

To the best of the author's knowledge, we have demonstrated for the first-time in literature the integration of a real-time neural network classifier processing directly the raw acoustic signal into the standard AudioMoth firmware, thereby proposing a Smart AudioMoth. The model is fully operational, runs in real-time and generates an increase of 18% in energy consumption for the continuous analysis of the soundscape. Our work validates a complete methodology from model design using 1D-CNNs on raw audio to on-device optimization and export with Aidge. The resulting system allows in-situ analysis and enables new functions with benefits in power and storage efficiency, paving the way for more effective and scalable PAM deployments. More advanced detection-based functions can easily be implemented upon our work such as alerts, selective compression or early interrupted recordings. This work also highlights the value of interdisciplinary collaboration between ecology surveying, machine learning and embedded systems engineering. Future directions include exploring on-device further optimization of the classification model and conducting in-situ studies using the smart AudioMoth system in real-world ecological projects.

## Web References

[1] Edge-impulse models on AudioMoth: https://github.com/OpenAcousticDevices/AudioMoth-EdgeImpulse

[2] Tiling and export tutorial: https://gitlab.eclipse.org/eclipse/aidge/aidge/-/blob/main/examples/tutorials/Tiling_1DCNN_tutorial/tiling.ipynb?ref_type=heads

[3] Demonstration video: https://youtu.be/7KY8f4xz0Z8?si=8-FyQHJRN9SYM420

## Acknowledgments

This work was supported in part by the DeepGreen Projet (ANR-23-DEGR-0001) and by INSITU chaire of MIAI (ANR-23-IACL-0006).
The authors would like to thank Aidge development group for their help, notably Pierre Gaillard and Olivier Bichler. We would also like to thank Valentin Baron, Maxime Bru, Maxime Peralta, Thibault Haltier, Olivier Antoni, Laurent Chiasson-Poirier, Diego Puschini, Yanis Basso-Bert and Paul Peyret for the various exchanges during this project.